\documentclass{PoS}

\title{On possible application of spin light of neutrino in astrophysics}

\ShortTitle{On possible application of spin light of neutrino in astrophysics}

\author{Alexander Grigoriev \\
       Skobeltsyn Institute of Nuclear Physics, Moscow State University, 119992 Moscow, Russia\\
       Department of Theoretical Physics, Moscow Institute of Physics and Technology, 141700 Dolgoprudny, Russia}

\author{Alexey Lokhov %\thanks{A footnote may follow.}
        \\
        Institute for Nuclear Research, Russian Academy of Sciences, 117312 Moscow, Russia}

\author{\speaker{Alexander Studenikin} \\
       Department of Theoretical Physics, Moscow State University, 119992 Moscow, Russia\\
       Dzhelepov Laboratory of Nuclear Problems, Joint Institute for Nuclear Research, 141980 Dubna, Russia \\
       E-mail: \email{studenik@srd.sinp.msu.ru}}

\author{Alexei Ternov \\
       Department of Theoretical Physics, Moscow Institute of Physics and Technology, 141700 Dolgoprudny, Russia}

\abstract{The \emph{spin light of neutrino} ($SL\nu$) is a phenomenon of electromagnetic radiation by a massive neutrino moving in external media that is originated owing to neutrino magnetic moment. In this short paper we note on the importance of this effect in the light of its connection with the neutrino magnetic moment, recap its basic properties in dense matter and give some general criteria for its best efficiency in nature. On this basis we propose a set of possible astrophysical environments where the $SL\nu$ can be manifested in principle.}

\FullConference{EPS-HEP 2017, European Physical Society conference on High Energy Physics\\
		5-12 July 2017\\
		Venice, Italy}

\begin{document}
Neutrino electromagnetic properties \cite{Giunti-Stud-RMP:2015, Studenikin:2017pos137} is the research field that is of fundamental significance and that also has important applications in astrophysics. Their existence is the natural consequence of nonzero neutrino mass which proof has found its final recognition recently \cite{Fukuda-Osc:98bar}--\cite{Ahmad-Osc:2002bar}. The most studied and understood among the neutrino electro\-magnetic characteristics are the dipole magnetic (diagonal, $i=j$, and transition, $i\neq j$) moments
\begin{equation}
\mu_{ij}=f_{M}{\mathstrut}_{ij}(0),
\end{equation}
given by the corresponding form factors  at $q^2=0$. The diagonal magnetic  moment of a Dirac neutrino in the minimally-extended Standard Model with right-handed neutrinos is \cite{Fuj-Shrock:80}
\begin{equation}\label{mu_D}
    \mu^{D}_{ii}
  = \frac{3e G_F m_{i}}{8\sqrt {2} \pi ^2}\approx 3.2\times 10^{-19}
  \Big(\frac{m_i}{1 \ \mathrm{eV} }\Big) \mu_{B},
\end{equation}
where $\mu_B$ is the Bohr magneton.
%This value is dramatically small, but there is a hope (see \cite{Giunti-Stud-RMP:2015}) that it can be much larger owing to contributions predicted in other generalizations of the Standard Model.

The best laboratory limit on neutrino magnetic moment has been obtained by the GEMMA collaboration from investigation of the reactor antineutrino-electron scattering at the Kalinin Nuclear Power Plant (Russia) \cite{GEMMA:2012}:
\begin{equation}\label{mu_bound}
\mu_{\nu} < 2.9 \times 10^{-11} \mu_{B} \ \ (90\% \ C.L.).
\end{equation}
The most recent stringent constraint on the electron neutrino effective magnetic moment
\begin{equation}
{\mu}_{\nu_e}\leq (2.8) \times
10^{-12} \mu _B
\end{equation}
has been reported by the Borexino Collaboration \cite{Borexino:2017fbd}.

The tightest astrophysical bounds are provided from the observed properties of globular cluster stars \cite{Raffelt-Clusters:90} (see also \cite{Viaux-clusterM5:2013,Arceo-Diaz-clust-omega:2015}) and from the time averaged neutrino signal
of SN1987A \cite{Kuznetsov:2009zm} and amounts, respectively, to
\begin{equation}
\mu_{\nu} \leq (2.2{-}2.6) \times
10^{-12} \mu _B, \ \ \ \bar{\mu}_{\nu}\leq (1.1{-}2.7) \times
10^{-12} \mu _B.
\end{equation}
The first constraint is applicable to both Dirac and Majorana neutrinos and the second one to Dirac neutrino.

The non-vanishing neutrino magnetic moment opens a realm of connected phenomena, of which the most discussing are the neutrino spin and spin-flavor oscillations (see \cite{Giunti-Stud-RMP:2015} and references therein). However the dramatic smallness of the neutrino magnetic moment makes direct studies of this important quantity extremely difficult. Nevertheless one can hope for a fruitful indirect study through a phenomenon which observational properties would depend on this quantity. Another conventional idea that can be invoked is to subject a neutrino to extremal conditions (e.g. strong magnetic fields and dense matter) in order to have the most strong manifestation of the neutrino properties. This proposition naturally connects investigation of the neutrino properties within astrophysical applications.

For a neutrino with nonzero magnetic moment there exists a potential capability of the direct coupling to a photon without change of the neutrino type
\begin{equation}\label{SLn}
\nu \rightarrow \nu +\gamma.
\end{equation}
This process being forbidden in vacuum can proceed under the external conditions. In \cite{Lobanov-Stud:03}, the process (\ref{SLn}) in external condition represented by the background matter was proposed. This new mechanism of electromagnetic radiation by a neutrino was termed ``the spin light of neutrino'' in matter. Within the classical treatment, the emission of light (photons) is due to the neutrino magnetic moment precession \cite{Lobanov-Stud:03,Lobanov-Stud:04}. In the quantum theory, the $SL\nu$ process is due to neutrino spin flip transition in matter \cite{Stud-Ternov-PLB:05bar}--\cite{Kuz-Mikh-anti:2007}.

The theory of the $SL\nu$ has been developed in above cited papers. In this short note we try to outline a set of astrophysical environments that are most suitable for $SL\nu$ applications . For this reason we review (see also \cite{Grigoriev:2017wff}) the basics of the $SL\nu$ and list the main $SL\nu$ properties to show which parameters are the most crucial for the radiation efficiency.

The neutrino spin light originates owing to the neutrino-photon magnetic moment coupling and the matter-induced splitting of neutrino energy levels over the helicity $s= \pm 1$. The last circumstance means existence of the energy gap between neutrino states in matter. It is determined by the neutrino dispersion relation (for definiteness we consider the Dirac neutrino) \cite{Stud-Ternov-PLB:05bar}
\begin{equation}\label{E}
    E_{\nu}=\sqrt{(p-s\tilde{n})^{2}+m_{\nu}^{2}}+\tilde{n},
 \label{dispersion}
\end{equation}
where, for example, for electron neutrino and ordinary matter composed of electrons, protons and neutrons with correspondent number densities $n_e$, $n_p$ and $n_n$ the ``density parameter'' $\tilde{n}$ reads
\cite{Stud-Ternov-PLB:05bar}:
\begin{equation}\label{n_nu}
  \tilde{n}_{\nu_e}=\frac{G_F}{2\sqrt{2}}\Big(n_e(1+4\sin^2 \theta
_W)+n_p(1-4\sin^2 \theta _W)-n_n\Big).
\end{equation}
For the case of neutrino moving through the matter composed of other neutrinos and antineutrinos
\begin{equation}
\tilde{n}_{\nu_{l^{\mathstrut}}\nu_{l^{\mathstrut\prime}}}=\frac{1}{2\sqrt{2}}{G}_{F}
(n_{\nu_{l^{\prime}}}-n_{\bar{\nu}_{l^{\prime}}}) \left(1+\delta_{ll^{\prime}}\right)  , \label{n_in_nu}
\end{equation}
where index $l$ refers to propagating neutrino flavor, index $l'$ refers to the matter neutrino flavor, and $n_{\nu_{l^{\mathstrut\prime}}}$ ($n_{\bar{\nu}_{l^{\mathstrut\prime}}}$) is the neutrino (antineutrino) number  density.

The $SL\nu$ photon (which in the general case should be treated as plasmon) momentum is obtained from the kinematical conservation laws. In the massless neutrino limit is has a simple expression
\begin{equation}\label{k_two-valued}
  k=p\ \frac{2\tilde{n}\cos\theta \pm \sqrt{4\tilde{n}^2-m_{\gamma}^2\sin^2\theta}}{4\tilde{n}+p(1-\cos ^2\theta)},
\end{equation}
where $\theta$ is the angle between the plasmon and initial neutrino momenta and $m_{\gamma}$ is the plasmon mass. The main features of the radiation connected with nonzero plasmon mass is the existence of the threshold which can be represented through the condition
\begin{equation}\label{Threshold}
  (m^2_{\gamma}+ 2\, m_{\gamma}m_{\nu})/4\tilde{n}p <1,
\end{equation}
and the two-valued dependance of the plasmon energy on the direction of the emission, which is expressed in presence of the sign ``$\pm$'' in Eq. (\ref{k_two-valued}).

The closed expressions for the total transition rate and power  of the radiation are rather cumbersome \cite{Grig-Stu-Ternov-Major:2006}. That is why for the purpose of assessment of the effect it makes sense to consider these quantities only for specific relations among the parameters. Since the energy gap between neutrino initial and final states is as larger as the neutrino density is greater  
(see Eqs. (\ref{E}) and (\ref{n_nu})) and because the radiation energy is proportional to $p$ here we will discuss the case of relativistic neutrinos ($m_{\nu}/p\ll 1$) and ``far above-threshold'' regime when the left-hand side of Eq. (\ref{Threshold}) is much smaller than unity. Under these conditions the total rate and radiation power are given by, respectively:
\begin{equation}\label{Gamma_I_rel}
  {\Gamma}=4{\mu}^2\tilde{n}^2 p, \ \ \ I={}^4\!\!/_3 {\mu}^2\tilde{n}^2 p^2.
\end{equation}
It is interesting to note that the average photon energy in this case
\begin{equation}\label{averaged_omega}
  \left< \omega \right>  \equiv I/{\Gamma} = {}^1\!\!/_3 p \simeq {}^1\!\!/_3 E_{\nu},
\end{equation}
meaning that the photon carries away a considerable part of neutrino's energy.

The $SL\nu$ radiation has non-trivial polarization properties that depend on relation among the matter density and neutrino energy \cite{Stud-Ternov-PLB:05bar,Grig-Stud-Ternov-PLB:05}. Under the conditions described above the radiation has a total circular polarization. This feature can be important for experimental identification of the radiation from dense astrophysical media.

Summing up the above we conclude that in order to get over a smallness of the neutrino magnetic moment and for the most effective realization of the $SL\nu$ effect one needs as high as possible values of neutrino energy and matter density. In this connection we note that a highest value of the matter density possible is attributed to neutron stars in which the particles number density reaches values up to $10^{38}$~cm$^{-3}$ \cite{Weber-Book:99}. In hypothetical third family compact stars the matter number density is expected to be an order of magnitude larger \cite{Kurkela-eos-AJ:2014}. The highest value of the neutrino energy measured is provided by results of the IceCube collaboration that has reported the
detection of extraterrestrial high-energy neutrinos with energy of about $E_{\nu}\sim 10^{16} $~eV \cite{IceCube-AJ:2015bar}. In astrophysics, the fluxes of cosmogenic neutrinos with energies up to $E_{\nu\,\max}\simeq$ $10^{20}{-}10^{22}\ $~eV are discussed \cite{Beres-Zatsepin:69}.

On this basis we can propose the following astrophysical environments potentially suitable for $SL\nu$ radiation applications \cite{Grig-Lokh-Stud-Ternov:17}:
\begin{itemize}
  \item $SL\nu$ radiation in neutron and quark stars;
  \item $SL\nu$ radiation by energetic neutrino moving in dense neutrino matter of supernova near the neutrinosphere (number density up to $10^{32}$~cm$^{-3}$ \cite{Qian-Woosley-wind:96});
  \item $SL\nu$ radiation during the development of gamma-ray burst (GRB) (neutrino number density up to $10^{32}$~cm$^{-3}$ for the model of short GRB \cite{Perego-Rosswog-etal-sRGB:2014}, neutron number density up to $10^{26}$~cm$^{-3}$ in the neutron-rich jet in the model of long GRB \cite{Vurm-Beloborodov-Poutanen-neutronGRB:2011});
  \item $SL\nu$ radiation by UHE-neutrinos moving through the relic neutrino background.
\end{itemize}
The third item contain the most interesting possibility since GRBs are considered as plausible sites for production of UHE-neutrinos \cite{Kumar-Zhang-GRB-rep:2015}. Moreover, due to its polarization properties in dense matter, the $SL\nu$ radiation may contribute to an account for the recently established problem of GRB radiation polarization.

It should be noted that the above listed cases can meet restrictions from the process threshold and competing processes. For example, in order to have the low threshold value for neutrino momentum one has to choose a matter with low charged component density. This circumstance makes it necessary to perform more rigorous estimations relevant to the listed cases to assess their feasibility and figure out which one is the most promising.

\vspace{5pt}

This work was supported by the Russian Foundation for Basic Research under grants
No.~16-02-01023\,A and No.~17-52-53133\,GFEN\_a.


\begin{thebibliography}{99}
\bibitem{Giunti-Stud-RMP:2015}
C.~Giunti and A.~Studenikin, %\emph{Neutrino electromagnetic interactions: a  window to new physics},
{\emph{Rev. Mod. Phys.} {\bf 87} (2015) 531}.

%\cite{Studenikin:2017pos137}
\bibitem{Studenikin:2017pos137}
  A.~Studenikin,
  %``Neutrino electromagnetic properties: a window to
%new physics - II,''
  PoS (EPS-HEP 2017)  137.
  %[arXiv:1706.01100 [hep-ph]].
  %%CITATION = ARXIV:1706.01100;%%

\bibitem{Fukuda-Osc:98bar}
{\scshape Super-Kamiokande} collaboration, Y.~Fukuda et~al., %\emph{{Evidence for oscillations of atmospheric neutrinos}},
{\emph{Phys. Rev. Lett.} {\bf
  81} (1998) 1562}.

\bibitem{Ahmad-Osc:2001bar}
{\scshape SNO} collaboration, Q.~R. Ahmad et~al., %\emph{{Measurement of the rate of $\nu_{e} + d \rightarrow p + p + e^{-}$ interactions produced by
  %${}^{8}\mathrm{B}$ solar neutrinos at the Sudbury Neutrino Observatory}},
  {\emph{Phys. Rev. Lett.} {\bf 87} (2001) 071301}.

\bibitem{Ahmad-Osc:2002bar}
{\scshape SNO} collaboration, Q.~R. Ahmad et~al., %\emph{{Direct evidence for
%  neutrino flavor transformation from neutral-current interactions in the
%  Sudbury Neutrino Observatory}},
  {\emph{Phys. Rev. Lett.} {\bf 89} (2002) 011301}.

\bibitem{Fuj-Shrock:80}
K.~Fujikawa and R.~E. Shrock, %\emph{{Magnetic moment of a massive neutrino and neu\-trino-spin rotation}},
  {\emph{Phys. Rev. Lett.} {\bf 45} (1980) 963}.

\bibitem{GEMMA:2012}
A.~G. Beda, V.~B. Brudanin, V.~G. Egorov, D.~V. Medvedev, V.~S. Pogosov, M.~V. Shirchenko et~al.,
  %\emph{{The results of search for the neutrino magnetic moment in GEMMA experiment}},
  {\emph{Adv. High Energy Phys.} {\bf 2012} (2012) 350150}.

\bibitem{Borexino:2017fbd}
  M.~Agostini {\it et al.} [Borexino Collaboration],
  %\emph{{Limiting neutrino magnetic moments with Borexino Phase-II solar neutrino data}},
  arXiv:1707.09355 [hep-ex].


\bibitem{Raffelt-Clusters:90}
G.~G. Raffelt, %\emph{{New bound on neutrino dipole moments from globular-cluster stars}},
   {\emph{Phys. Rev. Lett.} {\bf 64} (1990) 2856}.

  \bibitem{Viaux-clusterM5:2013}
N.~Viaux, M.~Catelan, P.~B. Stetson, G.~G. Raffelt, J.~Redondo, A.~A.~R. Valcarce et~al.,
   %\emph{Particle-physics constraints from the globular cluster {M5}: neutrino dipole moments},
   {\emph{Astron. \& Astrophys.} {\bf 558} (2013) A12}.

\bibitem{Arceo-Diaz-clust-omega:2015}
S.~Arceo-D\'{i}az, K.-P. Schr\"{o}der, K.~Zuber and D.~Jack,
  %\emph{{Constraint on the magnetic dipole moment of neutrinos by the tip-RGB luminosity in $\omega$-Centauri}},
  {\emph{Astropart. Phys.} {\bf 70} (2015) 1}.

\bibitem{Kuznetsov:2009zm}
  A.~V.~Kuznetsov, N.~V.~Mikheev and A.~A.~Okrugin,
  %\emph{{Reexamination of a Bound on the Dirac Neutrino Magnetic Moment from the Supernova Neutrino Luminosity}},
  {\emph{Int.\ J.\ Mod.\ Phys.\ A} {\bf 24} (2009) 5977}.

\bibitem{Lobanov-Stud:03}
A.~E. Lobanov and A.~I. Studenikin, %\emph{Spin light of neutrino in matter and electromagnetic fields},
  {\emph{Phys. Lett. B} {\bf 564} (2003) 27}.

\bibitem{Lobanov-Stud:04}
A.~Lobanov and A.~Studenikin, %\emph{Neutrino self-polarization effect in matter},
  {\emph{Phys. Lett. B} {\bf 601} (2004) 171}.

\bibitem{Stud-Ternov-PLB:05bar}
A.~I. Studenikin and A.~I. Ternov, %\emph{Neutrino quantum states and spin light in matter},
  {\emph{Phys. Lett. B} {\bf 608} (2005) 107}; [hep-ph/0410297].

\bibitem{Grig-Stud-Ternov-PLB:05}
A.~V. Grigoriev, A.~I. Studenikin and A.~I. Ternov,
  %\emph{Quantum theory of neutrino spin light in dense matter},
  {\emph{Phys. Lett. B} {\bf 622} (2005) 199}.

\bibitem{Grig-Stu-Ternov-Major:2006}
A.~V. Grigoriev, A.~I. Studenikin and A.~I. Ternov,
  %\emph{Dirac and {M}ajorana neutrinos in matter},
  {\emph{Phys. Atom. Nucl.} {\bf 69} (2006) 1940}.

\bibitem{Lobanov-SLnu:05bar}
A.~E. Lobanov,
  %\emph{High energy neutrino spin light},
  {\emph{Phys. Lett. B} {\bf 619} (2005) 136}.

\bibitem{Kuz-Mikh-anti:2007}
A.~V. Kuznetsov and N.~V. Mikheev,
  %\emph{{Plasma induced fermion spin-flip conversion $f_{L}\rightarrow f_{R} + \gamma$}},
  {\emph{Int. J. Mod. Phys. A} {\bf 22} (2007) 3211}.
%\cite{Grigoriev:2017wff}
\bibitem{Grigoriev:2017wff}
  A.~Grigoriev, A.~Lokhov, A.~Studenikin and A.~Ternov,
  %``Spin light of neutrino in astrophysical environments,''
  JCAP {\bf 1711} (2017) no.11,  024
  doi:10.1088/1475-7516/2017/11/024
  [arXiv:1705.07481 [hep-ph]].
  %%CITATION = doi:10.1088/1475-7516/2017/11/024;%%
\bibitem{Weber-Book:99}
F.~Weber,
  %\emph{{Pulsars as Astrophysical Laboratories for Nuclear and Particle Physics}}.
\newblock (Studies in High Energy Physics, Cosmology and Gravitation). IOP
  Publishing Ltd, Bristol, UK, 1999.

\bibitem{Kurkela-eos-AJ:2014}
A.~Kurkela, E.~S. Fraga, J.~Schaffner-Bielich and A.~Vuorinen,
  %\emph{Constraining neutron star matter with quantum chromodynamics},
  {\emph{Astrophys. J.} {\bf 789} (2014) 127}.

\bibitem{IceCube-AJ:2015bar}
{\scshape IceCube} collaboration, M.~G. Aartsen et~al.,
  %\emph{{A combined   maximum-likelihood analysis of the high-energy astrophysical neutrino flux
  %measured with IceCube}},
  {\emph{Astrophys. J.} {\bf 809} (2015) 98}.

\bibitem{Beres-Zatsepin:69}
V.~S. Beresinsky and G.~T. Zatsepin, %\emph{Cosmic rays at ultra high energies (neutrino?)},
  {\emph{Phys. Lett. B} {\bf 28} (1969) 423}.

\bibitem{Grig-Lokh-Stud-Ternov:17}
A. Grigoriev, A. Lokhov, A. Studenikin and A. Ternov,
  %\emph{Quantum theory of neutrino spin light in dense matter},
  arXiv:1705.0748 [hep-ph].

\bibitem{Qian-Woosley-wind:96}
Y.-Z. Qian and S.~E. Woosley, %\emph{Nucleosynthesis in neutrino-driven winds. {I.} {T}he physical conditions},
  {\emph{Astrophys. J.} {\bf 471} (1996) 331}.

\bibitem{Perego-Rosswog-etal-sRGB:2014}
A. Perego at al, %\emph{Neutrino-driven winds from neutron star merger remnants},
  {\emph{Mon. Not. R. Astron. Soc.} {\bf 443} 4 (2014) 3134}.

\bibitem{Vurm-Beloborodov-Poutanen-neutronGRB:2011}
I. Vurm, A. Beloborodov and J. Poutanen, %\emph{Gamma-ray bursts from magnetized collisionally-heated jets},
  {\emph{Astrophys. J.} {\bf 443} 4 (2011) 77}.

\bibitem{Kumar-Zhang-GRB-rep:2015}
P.~Kumar and B.~Zhang, %\emph{The physics of Gamma-Ray Bursts \& relativistic jets},
  {\emph{Phys. Rep.} {\bf 738} (2015) 1}.

\end{thebibliography}
\end{document}